# Magnetic behavior of spin-chain compounds, $Sr_3ZnRhO_6$ and $Ca_3NiMnO_6$, from heat capacity and ac susceptibility studies

**S. Rayaprol, Kausik Sengupta and E.V. Sampathkumaran**

*Tata Institute of Fundamental Research, Homi Bhabha Road, Colaba*

*Mumbai - 400005, India*

*And*

**Y. Matsushita**

*The Institute for Solid State Physics, The University of Tokyo, 5-1-5, Kashiwanoha,*

*Chiba 277-8581, Japan.*

## ABSTRACT

Heat-capacity (C) and ac susceptibility measurements have been performed on the spin-chain compounds, $Sr_3ZnRhO_6$ and $Ca_3NiMnO_6$, to establish their magnetic behavior and to explore whether there are magnetic frustration effects due to antiferromagnetic coupling of the chains arranged in a triangular fashion. While the paramagnetic Curie temperatures have been known to be large with a negative sign, as though antiferromagnetic interaction is very strong, the results establish that (i) the former apparently undergoes inhomogeneous magnetic ordering only around 15 K, however without spin-glass anomalies, and (ii) the latter orders antiferromagnetically at a relatively low temperature (17 K). Thus, the magnetic frustration manifests differently in these compounds.

**Corresponding author:** E.V. Sampathkumaran

E-mail address: sampath@tifr.res.in

Fax: 0091 22 2280 4610; phone: 0091 22 2280 4545

**PACS numbers:** 75.50. -y; 75.40.-s

**Key words:** Spin-chain compounds**,** $Sr_3ZnRhO_6$, $Ca_3NiMnO_6$, Heat-capacity.

# INTRODUCTION

The investigation of magnetic frustration effects due to topological reasons is one of the current topics in magnetism. In this regard, the compounds of the type, (Ca, Sr)$_3$ABO$_6$, crystallizing in the $K_4CdCl_6$-derived rhombohedral structure are of special interest, considering that the A and B ions present in the chains (along c-direction) are arranged in a triangular fashion in the basal plane [1,2]. Therefore, if A and/or B are magnetic ions, one would expect magnetic frustration effects, in the event that the interchain interaction is antiferromagnetic as it appears to be the case in this family of compounds. In fact, the compounds, $Ca_3CoRhO_6$ [3], $Ca_3Co_2O_6$ [4], $Ca_3CoIrO_6$ [5], and $Sr_3CuIrO_6$ [6] have been found to exhibit novel magnetic properties due to this topological frustration effects. However, the magnetic frustration does not appear to be a common rule for this class of compounds, as many other compounds (e.g., $Sr_3ZnIrO_6$ [6], $Ca_3CuMnO_6$ [7], $Ca_3CoMnO_6$ [8]) have been established to undergo long-range magnetic ordering, which may imply that the magnetic coupling with second (and higher) nearest neighbors play a vital role in deciding magnetic behavior. It is even more interesting to note that the creation of chemical and/or bond disorder in some stoichiometric spin-glass-like systems in this family favors long-range ordering [See Rayaprol et al in Ref. 4 and Sengupta et al in Ref. 7), whereas these factors should have favored spin-glass freezing from the common knowledge in magnetism. It is therefore of interest to subject other compounds in this family to careful investigations. In this article, we focus on two compounds, $Sr_3ZnRhO_6$ [9] and $Ca_3NiMnO_6$ [10]. The former compound is especially of interest from the chemist's point of view, as Zn is in a highly unusual trigonal prismatic coordination environment and also it forms in 'commensurate' as well as in 'incommensurate' structures [9], depending upon the methods of synthesis. Though certain inferences from the dc $\chi$ data with respect to magnetic ordering phenomena have been drawn earlier [9,10], dc $\chi$ results sometimes are not conclusive enough in this regard for this family of spin-chain compounds [11]. Therefore, we carried out heat capacity (C) and ac susceptibility ($\chi$) studies on these two compounds, the results of which are presented in this article.. We also discuss the results of isothermal magnetization (M) studies up to 120 kOe (for the first time for these compounds extended to such high fields).

## EXPERIMENTAL DETAILS

The compounds were prepared by the solid-state route. In the case of Zn sample, it is known [9] that this route results in the commensurate structure. For this compound, requisite amounts of high purity (>99.9%) $SrCO_3$, ZnO and Rh powders were mixed under acetone and preheated at $800^0$C for 10 hours. This powder was then pelletized and sintered at $1150^0$C for 9 days with three intermediate grindings. $Ca_3NiMnO_6$ was prepared by using high purity (>99.9%) $CaCO_3$, NiO and $MnO_2$ as starting components. A mixture containing these in proper proportion was calcined twice at $950^0$C for 24 hours each with an intermediate grinding and finally sintered in pellet form at $1250^0$C for 36 hours. The samples were characterized to be single phase by x-ray diffraction (Cu $K_\alpha$). There is no evidence for a secondary phase within the detection limit (<2%) of our x-ray diffraction (see Fig. 1) and the pattern obtained (including the relative intensities of the diffraction lines), for instance, for the Zn compound, is found to be in excellent agreement with that shown in Ref. 9. We have also performed Rietveld analysis for both the compounds and the results of such a refinement along with lattice parameters are shown in figure 1; such an analysis also establishes the formation of the Zn sample in the commensurate structure. We have also attempted to estimate oxygen content by thermogravimetric (TGA) analyses. Such studies in the case of Ca sample reveal that the stoichiometry of oxygen is close to 6.03; however, we could not precisely estimate the corresponding value for the Sr sample due to the continuous loss of ZnO beyond 850 C during TGA measurements; if one assumes that there is no loss of Zn at high temperatures, we can place an upper limit for the oxygen content at 6.2, which implies a value far below that expected for incommensurate structure (Ref. 9). Isothermal dc M measurements were performed up to 120 kOe employing a vibrating sample magnetometer (Oxford Instruments). Ac $\chi$ measurements (2 – 40 K) were carried out at four frequencies (1.3, 13, 130, 1330 Hz). The C measurements were performed by a semi-adiabatic heat-pulse method in the temperature (T) interval 2 – 40 K. In addition, dc $\chi$ measurements in the presence of 5 kOe have been carried out to compare with previous reports.

## RESULTS AND DISCUSSION

The results of magnetization and heat-capacity measurements are shown in Fig. 2 and Fig. 3 respectively for both the samples.

In the case of Zn sample, $\chi$ follows Curie-Weiss behavior above 50 K, below which there is a deviation from high-T linearity, as noted by Layland and Loye [9]. The values of the effective moment ($\mu_{eff}$ = 2.12$\mu_B$) and the magnitude of the paramagnetic Curie temperature ($\theta_p$ = -100 K) are marginally larger than those reported by these authors for the commensurate structure (1.64 $\mu_B$ and -70 K respectively). This difference in the value of $\mu_{eff}$ implies some degree of sample dependence, which could be associated with strong anisotropic nature of the magnetic interaction in such spin-chain compounds. This makes the situation to infer the valence/spin state of Rh very difficult from the magnetization data alone. However,

Rh valence should be 4, considering that other elements take a fixed value for the valence. The large value of $\theta_p$ with a negative sign observed in the present studies as well as in Ref. 9 implies strong antiferromagnetic correlations as though this compound may order antiferromagnetically at rather high temperatures. In contrast to this expectation, it is reported in Ref. 9 that there is no evidence for *long range* ferro/antiferro magnetic ordering down to 2 K. While the results presented in this article (see below for further discussions) render an overall support to this conclusion, we see a broad peak around 15-20 K both in the high field (5 kOe) as well as in the low-field $\chi$ (100 Oe) data. [In addition, there is an upturn in $\chi$ below 10 K, which will be addressed at the end of this paragraph]. The isothermal M curves taken up to 120 kOe at 5 and 15 K are found to be essentially linear (with similar values at both the temperatures), which establishes that the compound is not ferromagnetic down to 2 K. In order to understand the origin of this feature in $\chi$ further, we have performed C measurements, the results of which are shown in Fig. 3. It is to be to be noted that there is no evidence for any prominent $\lambda$-anomaly, expected for long range magnetic ordering, in the entire T-range of magnetic ordering. However, there is a broad shoulder around 15 K, which is more clearly visible in the plot of C/T (see inset of figure 3a). The presence of this shoulder implies the existence of *short range* (or inhomogeneous) magnetic ordering or spin-glass freezing. We have carefully performed ac $\chi$ measurements and we do not find any peak both in real ($\chi$') as well as imaginary ($\chi$") parts, in the range 2 to 60 K at any of the frequencies (and hence not shown in the form of a figure). This establishes that there is no spin-glass freezing in this material. In support of this finding, the zero-field-cooled (ZFC) and field-cooled (FC) dc $\chi$. curves obtained at a low field (100 Oe) follow each other without significant bifurcation as mentioned in Ref. 9. Thus, the broad 15K-anomaly in C and a corresponding broad $\chi$ feature arise from inhomogeneous magnetic order. The broad peak in $\chi$ was not clearly resolved in Ref. 9 presumably due to masking by the large low temperature tail arising from paramagnetic impurities. We now turn to this low temperature (below 10 K) upturn in $\chi$. The fact that the values of $\chi$ at 100 Oe and 5 kOe, say, at 2 K, are comparable, and that M value even at very high H values (see Fig. 2c) are practically the same at 5 and 15 K without any evidence for high-field saturation establish that this upturn could be due to traces of paramagnetic impurity. This conclusion gains further credence from the fact that the magnitude of the tail below 10 K as observed by us is much smaller than that seen in Ref. 9, which should not be the case if the upturn is intrinsic to this material. In short, all these results reveal that there is inhomogeneous magnetic ordering around 15 K for this compound, without any further long range ordering at lower temperatures.

We now address the magnetic behavior of $Ca_3NiMnO_6$. As far as $\chi$ plot is concerned, there is an overall agreement with that reported in Ref. 10. We observe a peak due to magnetic ordering at 17 K. Curie-Weiss law is obeyed in the range 100 to 300 K and the $\mu_{eff}$ and $\theta_p$ in this range turn out to be 6 $\mu_B$ and -380 K respectively. The value of $\mu_{eff}$ is larger than that reported (5.1 $\mu_B$) in Ref. 10, and a possible origin of this discrepancy was addressed above. It is interesting that the magnetic ordering takes place at a much lower temperature compared to the value of $\theta_p$, as though there are competing interactions leading to magnetic frustration effects. In order to address whether this magnetic frustration leads to spin-glass freezing, careful C and ac $\chi$ measurements have been carried out. With respect to the C data (Fig. 3b), there is a well-defined sharp peak at 17 K, characteristic of long range ordering, but not of spin-glass freezing. In support of this, we do not see significant bifurcation of ZFC-FC $\chi$ curves. We see a change in the slope in the plot of real part of ac $\chi$ around 17 K, however, without any frequency dependence of this temperature; in addition, the imaginary part is featureless in the entire T range of investigation; these findings offer strong evidence against spin-glass freezing. At 5K, M is found to vary nearly linearly with T without any hysteresis at low fields, which is actually characteristic of antiferromagnets. These results viewed together establish that this compound undergoes long range antiferromagnetic ordering below 17 K, supporting the conclusion from the neutron diffraction data [10]. Finally, we note that the broad peak in $\chi$ at about 100 K noted in Ref. 10 in favor of one-dimensional magnetism is not observable in our data.

## CONCLUSION

To conclude, present heat-capacity and ac susceptibility results reveal inhomogeneous magnetic order around 15 K in the compound, $Sr_3ZnRhO_6$, however without spin-glass features, while in $Ca_3NiMnO_6$ long range antiferromagnetic order is established to develop at a rather low temperature (17 K) compared to the magnitude of the paramagnetic Curie temperature. Clearly, geometrical frustration phenomenon plays a role in deciding magnetism of these compounds.

The authors would like to thank Kartik K Iyer for his help during measurements.

REFERENCES

1. See for instance, T.N. Nguyen and H.C. zur Loye, J. Solid State Chem. **117,** 300 (1995).
2. K.E. Stitzer, W.H. Hemley, J.B. Claridge, H.C. zur Loye and R.C. Layland, J. Solid State Chem., **164,** 220 (2002) and references therein.
3. S. Niitaka, K. Yoshimura, K. Kosuge, M. Nishi, and K. Kakurai, Phys. Rev. Lett. 87, 177202 (2001); E.V. Sampathkumaran and Asad Niazi, Phys. Rev. B **65**, 180401(R) 2002.
4. H. Kageyama, K. Yoshimura, K. Kosuge, H. Mitamura and T. Goto, J. Phys. Soc. Japan. **66**, 1607 (1997); A. Maignan, A.C. Masset, C. Martin, and B. Raveau, Eur. Phys. J. B **15**, 657 (2000); S. Rayaprol, K. Sengupta and E.V. Sampathkumaran, Solid State Commun. **128**, 79 (2003); S. Rayaprol, K. Sengupta and E.V. Sampathkumaran, Proc. Indian. Acad. Sci. (Chem. Sci), **115**, 553 (2003).
5. S. Rayaprol, Kausik Sengupta, and E.V. Sampathkumaran, Phys. Rev. B 67, 180404(R) (2003).
6. Asad Niazi, P.L. Paulose and E.V. Sampathkumaran, Phys. Rev. Lett. **88,** 107202 (2002); Asad Niazi, P.L. Paulose, E.V. Sampathkumaran, D. Eckert, A. Handstein, and K.-H. Müller, Phys. Rev. B **65**, 064418 (2002).
7. Kausik Sengupta, S. Rayaprol, Kartik K Iyer, and E.V. Sampathkumaran, Phys. Rev. B **68**, 012411 (2003).
8. V.G. Zubkov, G.V. Bazuev, A.P. Tyutyunnik, and I.F. Berger, J. Solid State Chem. **160**, 293 (2001); also see, Rayaprol et al in Ref. 4.
9. R.C. Layland, and H.C. Zur Loye, J. Alloys and Compounds. **299,** 118 (2000).
10. S. Kawasaki, M. Takano, T. Inami, J. Solid State Chem. **145,** 302 (1999).

11. See, for instance, Asad Niazi, E.V. Sampathkumaran, P.L. Paulose, D. Eckert, A. Handstein and K.-H. Müller, Solid State Commun. **120**, 11 (2001).

Figure Captions

Fig.1.

Observed (+) and calculated (solid line) X-ray diffraction patterns of $Sr_3ZnRhO_6$ and Ca3NiMnO6. The positions of allowed Bragg reflections are shown by vertical ticks. The difference between calculated and observed lines is shown at the bottom of respective figures. The lattice constants and least squares refinement results (for commensurate structure in the case of the former) are also given and the notations are the same as those used in Ref. 9.

Fig. 2.

Magnetic susceptibility ($\chi$) and inverse $\chi$ as a function of temperature taken in a field of 5 kOe for (a) $Sr_3ZnRhO_6$ and (b) $Ca_3NiMnO_6$. The data obtained in a field of 100 Oe for zero-field-cooled (data points) and field-cooled (continuous line) conditions of the specimens are shown in (c) and (d) respectively and isothermal magnetization (M) data at selected temperatures (continuous line, 5 K; dotted line, 15, 30 K) are shown in the insets. The real part ($\chi$') of ac $\chi$ is plotted for 1.3 Hz for $Ca_3NiMnO_6$ as an inset in Fig. (d). In the case inverse $\chi$ plots, the continuous lines represent high-T linear region.

Fig. 3.

Heat capacity (C) as a function of temperature for (a) $Sr_3ZnRhO_6$ and (b) $Ca_3NiMnO_6$ in the range 2 - 40 K. In the inset of (a), C/T behavior is plotted and an arrow in (a) marks the broad shoulder discussed in the literature.

Figure 1

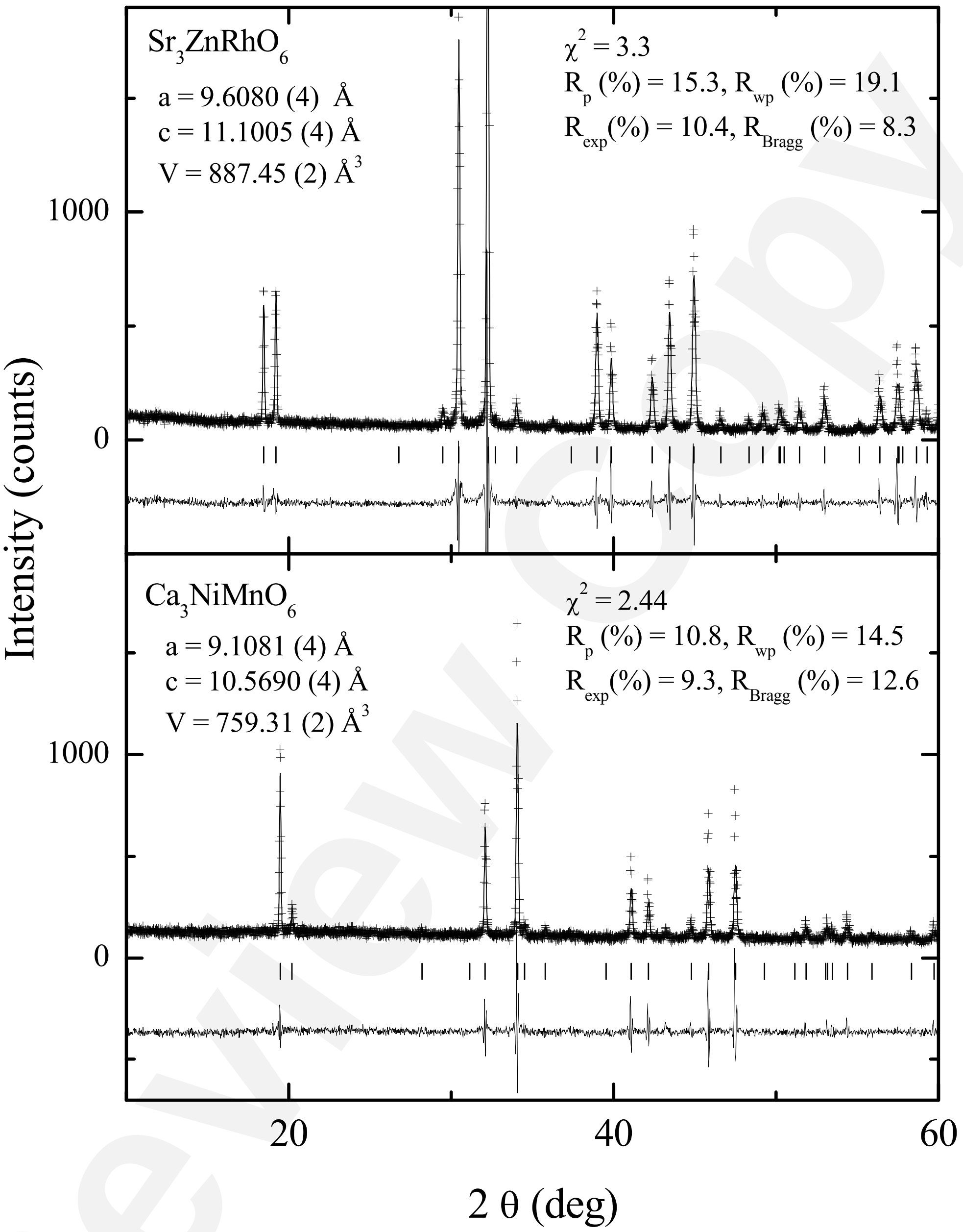
Sr3ZnRhO6
a = 9.6080 (4) Å
c = 11.1005 (4) Å
V = 887.45 (2) Å3
χ2 = 3.3
Rp (%) = 15.3, Rwp (%) = 19.1
Rexp(%) = 10.4, RBragg (%) = 8.3
Ca3NiMnO6
a = 9.1081 (4) Å
c = 10.5690 (4) Å
V = 759.31 (2) Å3
χ2 = 2.44
Rp (%) = 10.8, Rwp (%) = 14.5
Rexp(%) = 9.3, RBragg (%) = 12.6
Intensity (counts)
1000
0
20
40
60
2 θ (deg)

Figure 2

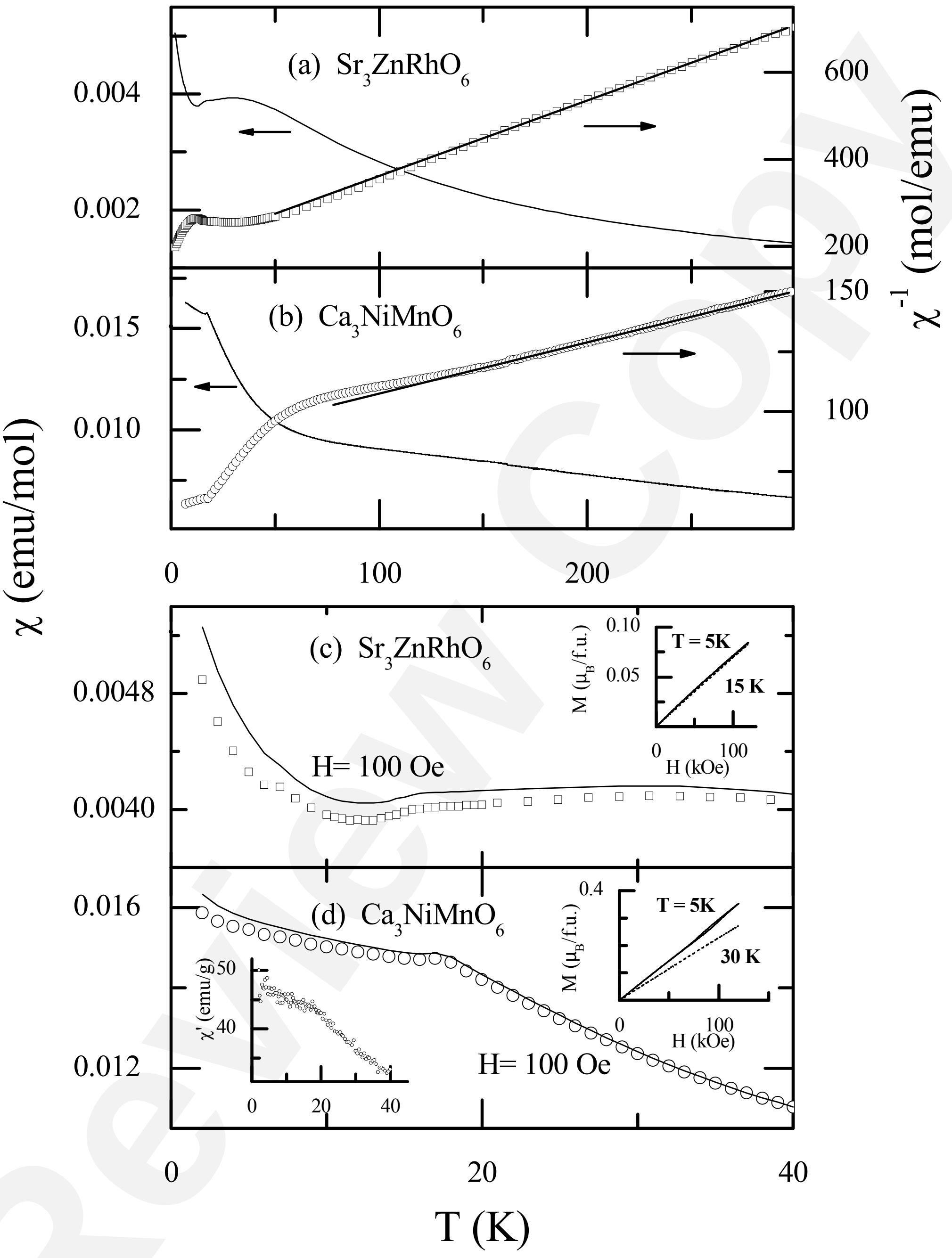

Figure 3

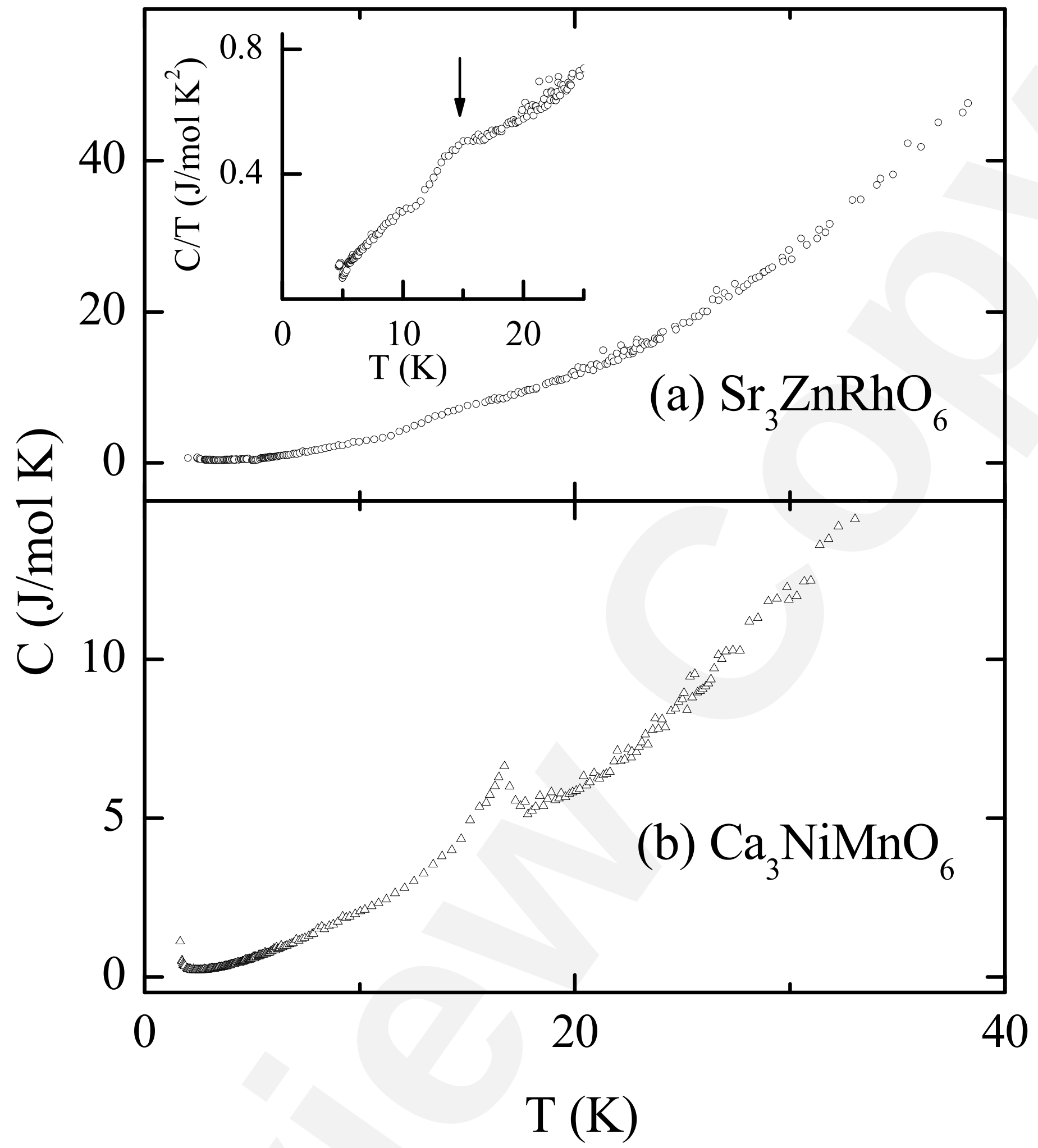